%% file: main.tex
\documentclass[preprint,twocolumn,12pt,5p]{elsarticle}
\usepackage{geometry}
\usepackage{amsmath,amssymb,subfig,booktabs,tikz,grffile,textgreek,tikz,multirow,circuitikz}
\journal{NIM A}
\begin{document}

\begin{frontmatter}

\title{A new beamline for laser spin-polarization at ISOLDE}

\author[iks]{W. Gins\corref{cor1}}
\ead{wouter.gins@fys.kuleuven.be}
\cortext[cor1]{Corresponding author}
\author[cern,york]{R. D. Harding}
\ead{robert.harding@cern.ch}
\author[poznan]{M. Baranowski}
\author[manchester]{M. L. Bissell}
\author[iks]{R. F. Garcia Ruiz\fnref{fn1}}
\fntext[fn1]{Current affiliation: EP Department, CERN, CH-1211 Geneva 23, Switzerland}
\author[cern]{M. Kowalska}
\author[iks,cern]{G. Neyens}
\author[cern]{S. Pallada}
\author[iks]{N. Severijns}
\author[iks]{Ph. Velten}
\author[cern]{F. Wienholtz}
\author[iks]{Z. Y. Xu}
\author[iks]{X. F. Yang}
\author[npi]{D. Zakoucky}

\address[iks]{KU Leuven, Instituut voor Kern- en Stralingsfysica, Leuven, Belgium}
\address[cern]{EP Department, CERN, Switzerland}
\address[york]{University of York, York, United Kingdom}
\address[poznan]{Faculty of Physics, Adam Mickiewicz University, Poznan, Poland}
\address[manchester]{School of Physics and Astronomy, The University of Manchester, Manchester, United Kingdom}
\address[poznantech]{Poznan University of Technology, Poznan, Poland}
\address[npi]{NPI, Czech Academy of Sciences, Rez, Czech Republic}

\begin{abstract}
A beamline dedicated to the production of laser-polarized radioactive beams has been constructed at ISOLDE, CERN. We present here different simulations leading to the design and construction of the setup, as well as technical details of the full setup and examples of the achieved polarizations for several radioisotopes. Beamline simulations show a good transmission through the entire line, in agreement with observations. Simulations of the induced nuclear spin-polarization as a function of atom-laser interaction length are presented for $^{26,28}$Na, \cite{0954-3899-44-8-084005} and for $^{35}$Ar, which is studied in this work. Adiabatic spin rotation of the spin-polarized ensemble of atoms, and how this influences the observed nuclear ensemble polarization, are also performed for the same nuclei. For $^{35}$Ar, we show that multiple-frequency pumping enhances the ensemble polarization by a factor 1.85, in agreement with predictions from a rate equations model.
\end{abstract}

\end{frontmatter}
\section{Introduction}

Spin-polarized radioactive nuclei have been a staple of nuclear and particle physics research since the discovery of parity violation \cite{PhysRev.105.1413}. With the use of polarized nuclei as a probe in fields ranging from fundamental interactions to material and life sciences \cite{rogar2015,Kowalska:2299798,Velten2014}, an initiative for a dedicated experiment at ISOLDE was started, and a beamline was built and commissioned. Results from the commissioning of the new beamline have been reported in Ref.~\cite{0954-3899-44-8-084005}. The present article documents the technical aspects of the beamline.

Section~\ref{sec:polarization} describes the mechanism of laser polarization through optical pumping and how the induced nuclear polarization can be observed through the asymmetry in the nuclear \textbeta-decay. Section~\ref{sec:beamline} reports on the different parts of the beamline, followed by Section~\ref{sec:beamsim} dedicated to the ion-optical simulations. The magnetic fields generated by the setup are discussed in Section~\ref{sec:fields}. The successful use of multiple-frequency optical pumping to achieve higher polarization for $^{35}$Ar atoms, to be used in future fundamental interactions studies \cite{Velten2014}, is described in Section~\ref{sec:multipump} and compared with experimental data. Calculations of the adiabatic rotation of the spin-polarized ensembles of $^{26,28}$Na and $^{35}$Ar in the magnetic fields are presented in Section~\ref{sec:adiabaticrotation} and compared to the observed asymmetries. Conclusions are given in Section~\ref{sec:conclusion}.

\section{Optical pumping and \textbeta-asymmetry}\label{sec:polarization}

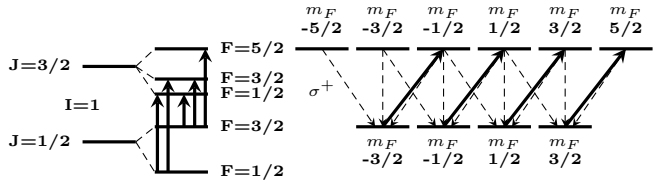
\begin{figure}[t!]
    \centering
    \input{pumping.txt}
\caption{Optical pumping scheme in the D2 line of $^{28}$Na. Excitations using $\sigma^+$ polarized light are drawn with solid lines, while the dashed lines indicate decay paths through photon emission.}
\label{fig:opticalpumping}
\end{figure}

The hyperfine interaction couples the nuclear spin $\vec{I}$ and the electron spin $\vec{J}$ together to a total atomic spin $\vec{F}=\vec{I}+\vec{J}$, which splits a fine structure level into several hyperfine levels. Atomic population can be resonantly transferred from one hyperfine level to a radiatively coupled level through interaction with a narrowband laser. In this process, conservation of angular momentum dictates that the atomic spin $\vec{F}$ can only change by maximally one unit. The left side of Figure~\ref{fig:opticalpumping} illustrates the 5 allowed hyperfine transitions for the D2 line in $^{28}$Na ($3^2S_{1/2}\rightarrow 3^2P{3/2}$) as solid lines. The term ``optical pumping'' is used when resonant excitations and decay drive the atomic population towards a specific (magnetic sub)state \cite{Kastler1957}.

\begin{figure*}[t!]
\centering
    \includegraphics[width=1\textwidth]{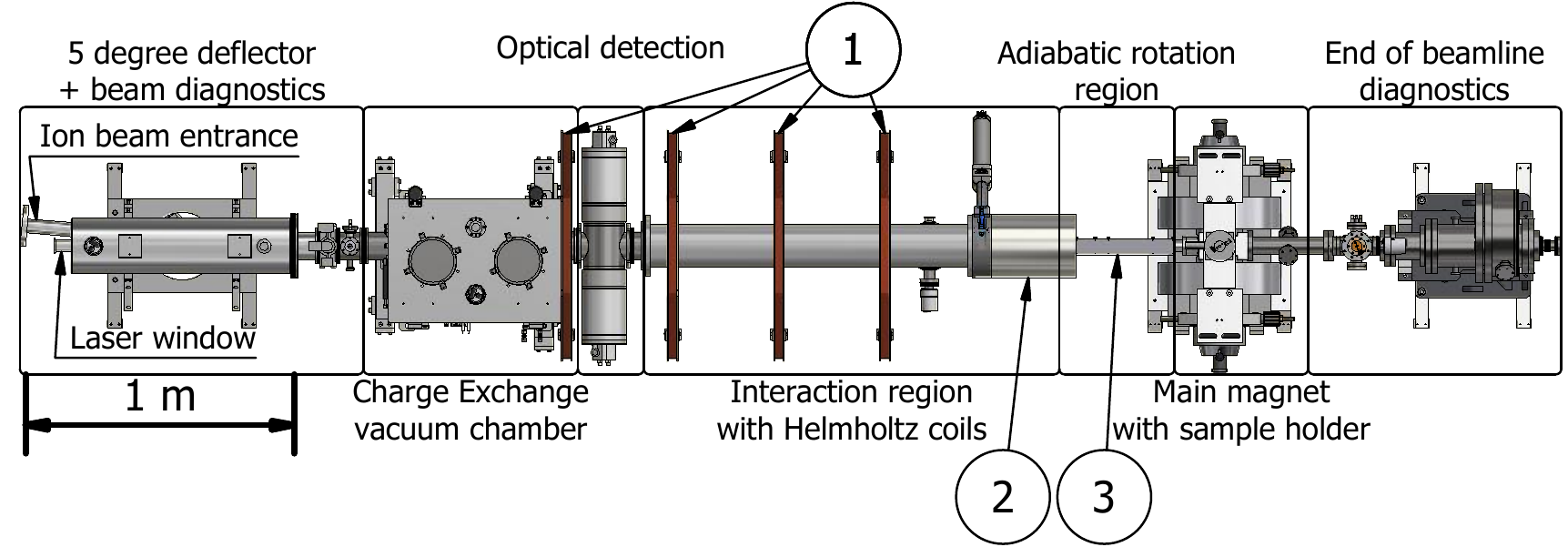}
    \caption{Schematic overview of the VITO beamline, with the main beamline components indicated. The beam from ISOLDE enters from the top left beampipe, and the start of this beampipe is used as the start for the ion-optical simulations. The numbers are used to indicate the coils in Table~\ref{tab:solenoids}.}
    \label{fig:beamline}
\end{figure*}

When the laser light is circularly polarized ($\sigma^+$ or $\sigma^-$), conservation of angular momentum further imposes the restriction $\Delta m_F=1$ or $-1$ (indicated as solid lines on the right side of Figure~\ref{fig:opticalpumping}). The decay back to the lower state is however not bound to this rule and can proceed with $\Delta m_F=0,\pm 1$ (indicated by dashed lines) \cite{Demtroder1981}. By maximizing the number of such excitation/decay cycles, the population of a specific F-state is pushed towards substates with either maximal or minimal $m_F$ quantum numbers in the lowest level.

The number of excitation/decay cycles can be increased by either increasing the laser photon density or by having a longer interaction time. In a collinear geometry (as used in high-resolution collinear laser spectroscopy experiments \cite{Neugart2017}), where the particle and CW laser beam are spatially overlapped, such an enhanced interaction time is achieved by choosing an appropriate length of the laser-particle interaction region.

The resulting atomic polarization is then transferred to a polarization of the nuclear spin through the hyperfine interaction. The nuclear polarization in such an ensemble of nuclei is defined as
\begin{equation}
P=\sum_{m_I}\frac{w\left(m_I\right)m_I}{I},\label{eq:pol}
\end{equation}
with $w\left(m_I\right)$ the probability that the $\left|I,m_I\right\rangle$ quantum state is populated after the optical pumping process.

The nuclear polarization can be observed by detecting the asymmetry in the \textbeta-decay of radioactive isotopes, due to the parity violation in nuclear \textbeta-decay \cite{PhysRev.105.1413}. The \textbeta-decay of a polarized ensemble has a specific angular distribution that can be reduced to $W\left(\theta\right) \sim 1+AP\cos\left(\theta\right)$ for allowed \textbeta-transitions \cite{Jackson1957}, where $\theta$ is the angle between the emitted electron momentum and the nuclear spin orientation. $A$ is the asymmetry parameter of the decay which depends on the initial and final spin of the nuclear states involved in the \textbeta-decay, and $P$ is the polarization of the nuclear ensemble with respect to a quantisation axis. The experimental asymmetry (as reported in Section~\ref{sec:multipump}) is then defined as
\begin{equation}
    A_{exp} = \frac{N\left(0^\circ\right)-N\left(180^\circ\right)}{N\left(0^\circ\right)+N\left(180^\circ\right)}=\epsilon AP,
\end{equation}
where $\epsilon$ represents depolarization effects which depend on several experimental factors.

For more details on spin-polarization via optical pumping, see Ref.~\cite{Yordanov2007,Gins2018b}.

\section{Beamline description}\label{sec:beamline}

The laser-polarization setup, part of the Versatile Ion Techniques Online (VITO) beamline at ISOLDE, CERN \cite{Stachura2016}, is designed to polarize nuclei through the application of optical pumping of an atomic hyperfine transition. More than 1000 radioactive isotopes of more than 70 elements can be produced at ISOLDE, CERN, via the impact of a 1.4 GeV proton beam on a variety of targets using different types of ion sources \cite{Kugler2000}. The ion beam is sent into the laser polarization beamline and overlapped with a circularly polarized laser beam to induce nuclear polarization. After implantation in a suitable host, placed between the poles of an electromagnet, the change in \textbeta-asymmetry is observed as a function of laser frequency, scanned across the hyperfine structure. Although the technique is also applicable to ions \cite{Kowalska2008}, the current design is specific for working with atoms.
An overview of the layout of the entire optical pumping beamline is given in Figure~\ref{fig:beamline}.

The first element of the beamline is a 5 degree deflector equipped with a laser window, where the laser beam is overlapped with the pulsed radioactive ion beam. A beam diagnostics box, containing an adjustable iris to define the beam waist and a readout plate for the ion-current, is placed directly after the deflector. Beamline simulations of the 5 degree deflector are discussed in Section~\ref{sec:beamsim}.

\begin{figure}[!ht]
    \centering
    \includegraphics[width=\columnwidth]{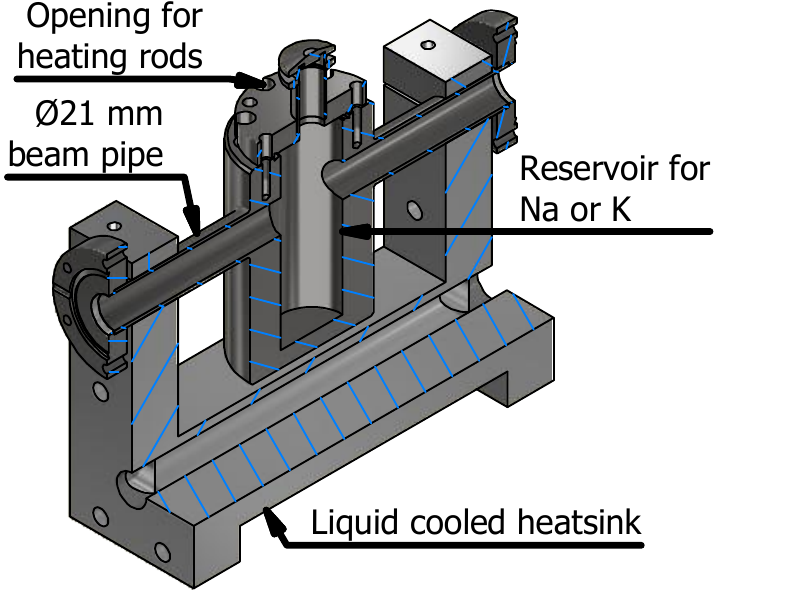}
    \caption{3/4 view of the CAD drawing of the CEC. The stainless steel reservoir has 6 deep holes to house heating rods. The liquid cooled heatsink clamps onto the beampipe ends.}
    \label{fig:cec}
\end{figure}

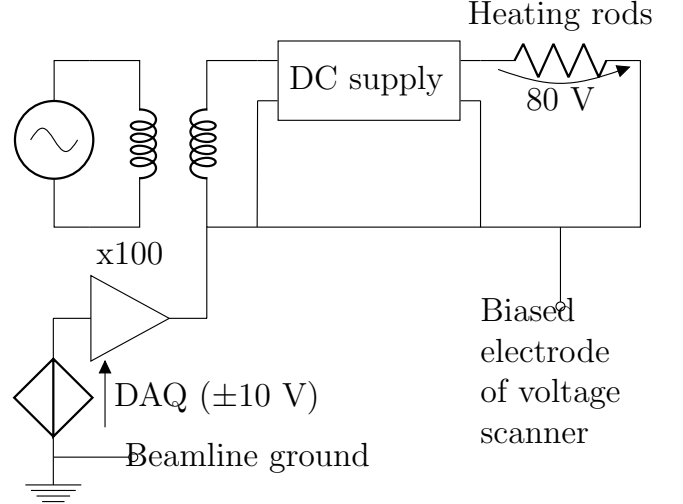
\begin{figure}[!ht]
    \centering
    \resizebox{\columnwidth}{!}{
    \begin{circuitikz}[scale=1.0]
        \draw (0, 0) node[oscillator](ac_source){};
        \draw (-0.5, -4) node[ground](dc_ground){};
        \draw (1, 1) node[transformer,box](transf){};
        \coordinate (transf_neutral) at ($(transf.B1)!0.535!(transf.B2)-(0.51,0)$);
        \coordinate (transf_neutral_east) at ($(transf_neutral)+(0.2,0)$);
        \coordinate (transf_bottom) at ($(transf.B2)!0.3!(transf.A2)$);
        \draw (dc_ground) to [cV_=DAQ ($\pm10$ V)] ++(0,1.5) to ++(0,0.25) to [amp,l=x100] (transf_bottom |- 0,-2.25) to (transf_bottom);
        \draw (ac_source.north) to (ac_source.north |- transf.A1) to (transf.A1);
        \draw (transf.A2) to (ac_source.south |- transf.A2) to (ac_source.south);
        \draw (transf.B2 |- ac_source)++(1.4,0.7465) node[draw,text width=2cm,minimum height=1cm](DC){DC supply};
        \coordinate (DCin1box) at ($(DC.north west)!0.25!(DC.south west)$);
        \coordinate (DCin1) at ($(DCin1box)+(-0.25,0)$);
        \draw (DCin1) to (DCin1box);
        \coordinate (DCin2box) at ($(DC.north west)!0.75!(DC.south west)$);
        \coordinate (DCin2) at ($(DCin2box)+(-0.25,0)$);
        \draw (DCin2) to (DCin2box);

        \coordinate (DCout1box) at ($(DC.north east)!0.25!(DC.south east)$);
        \coordinate (DCout1) at ($(DCout1box)+(0.25,0)$);
        \draw (DCout1) to (DCout1box);
        \coordinate (DCout2box) at ($(DC.north east)!0.75!(DC.south east)$);
        \coordinate (DCout2) at ($(DCout2box)+(0.25,0)$);
        \draw (DCout2) to (DCout2box);
        \coordinate (DCout1far) at ($(DCout1)+(2,0)$);

        \draw (transf.B1) to (DCin1);
        \draw (transf.B2) to (DCin2);
        \draw (DCout1) to[R,l=Heating rods,v=80 V] (DCout1far) to (DCout1far |- transf.B2) to (DCout2 |- transf.B2) to (DCout2);
        \draw (transf.B2) to (DCout2|- transf.B2);
        \draw (DCout2|- transf.B2)+(1,0) to[short,-o] node[anchor=north,text width=2cm]{Biased electrode of voltage scanner}++(1,-1);
        \draw (dc_ground) to[short,-o] node[anchor=west]{Beamline ground} ++(1,0);

    \end{circuitikz}
    }
    \caption{Simplified diagram of the electrical wiring for the voltage scanning. The voltage scanning provides a floating potential for a secondary circuit, where an isolating transformer supplies power to a DC power supply for the heating rods mentioned in Section~\ref{sec:beamline}. The data acquisition program can supply $\pm 10$V, which is amplified by a factor 100.}
    \label{fig:elecdiagram}
\end{figure}

The Charge Exchange Cell (CEC), housed in the Charge Exchange vacuum chamber of Figure~\ref{fig:beamline} and depicted in Figure~\ref{fig:cec}, is placed after the diagnostic box, where the beam passes through a vapor of Na or K and undergoes charge exchange. It contains a reservoir in the middle where solid Na or K is deposited. The stainless steel reservoir is heated using six RS-8607016 220 W heating cartridges powered by a DC power supply. To minimze the diffusion of the Na or K vapor to the rest of the beamline, the ends of the pipe are kept at a lower temperature. This is achieved using a heatsink cooled with circulating Galden\textsuperscript{\textregistered} PFPE down to 90$^\circ$ C, below the evaporation temperature of Na or K in vacuum. This minimizes the loss of vapor to the rest of the beamline and ensures a high vapor density in the middle. Non-neutralized ions are deflected from the beam after the CEC using an electrostatic deflector. A specially designed electrode arrangement (referred to as \emph{voltage scanner}, design details in Section~\ref{sec:beamsim}) is attached to this cell and modifies the kinetic energy of the incoming ion beam. This changes the velocity of the beam, and induces a Doppler shift of the laser frequency. The relation between the labframe $\nu_{rest}$ frequency and the frequency $\nu_{obs}$ observed by this accelerated (or decelerated) beam is
\begin{equation}
    \nu_{obs} = \nu_{rest}\sqrt{\frac{1-\beta}{1+\beta}},\label{eq:reldoppler}
\end{equation}
\begin{equation}
    \beta=\sqrt{1-\left(mc^2/\left(mc^2+qE_{kin}\right)\right)^2}.
\end{equation}
This allows a fast scanning of the hyperfine structure by changing the acceleration voltage. In Figure~\ref{fig:elecdiagram} we show the electrical diagram of the wiring that enables this voltage scanning. The data acquisition system (DAQ) can provide a controlled voltage of up to $\pm10$ V, which is amplified by a Kepco amplifier by a factor of 100. This voltage of $\pm1$ kV (and typical precision of 0.02 V) is then applied to the secondary windings of an isolating transformer. The insulating transformer supplies the 230 V line voltages on the secondary side to a DC power supply which is then also floated by the $\pm1$ kV. Since the biased electrode of the voltage scanner is connected to the base of the CEC and thus to the lower output of the power supply, both elements are biased to $\pm1$ kV relative to the beamline ground while a constant voltage is applied over the heating rods.

Following the CEC is the optical detection region, which is a copy of the light collection region used in the COLLAPS setup \cite{kreim2014}. The photomultiplier tubes in this detection area are used for determining the resonant laser frequency through optical detection of the fluorescence decay from a stable isotope of the element of interest, prior to starting \textbeta-asymmetry measurements on the less abundant radioactive species.

\begin{figure}[!ht]
\centering
    \includegraphics[width=\columnwidth]{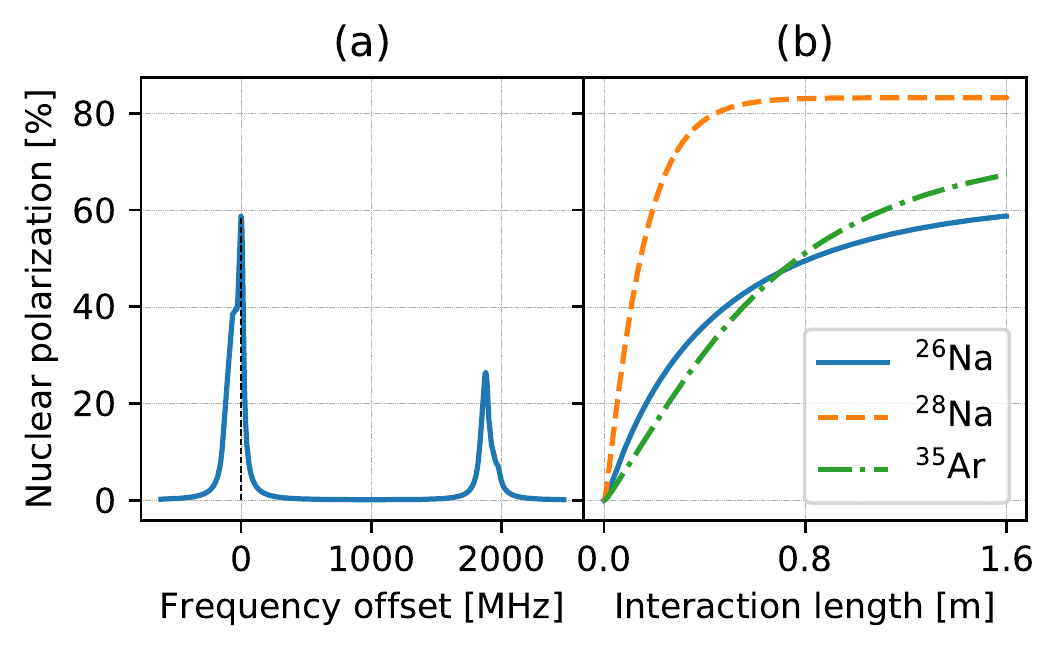}
    \caption{(a) Simulated hyperfine spectrum for pumping $^{26}$Na in the D2 line using $\sigma+$ polarized light at a typical laser power of 80 mW/cm$^2$. The interaction time corresponds to an interaction length of 1.6 m for a 50 keV beam. (b) The calculated polarization in the strongest component of the hyperfine spectrum (indicated with a dashed line in (a)) as a function of interaction length for the different nuclear species discussed in this paper.}
    \label{fig:designcalc}
\end{figure}

\begin{figure*}[!t]
\centering
    \begin{tikzpicture}[scale=0.90]
        \draw (0, 0) node[inner sep=0]{\includegraphics[width=0.7\textwidth]{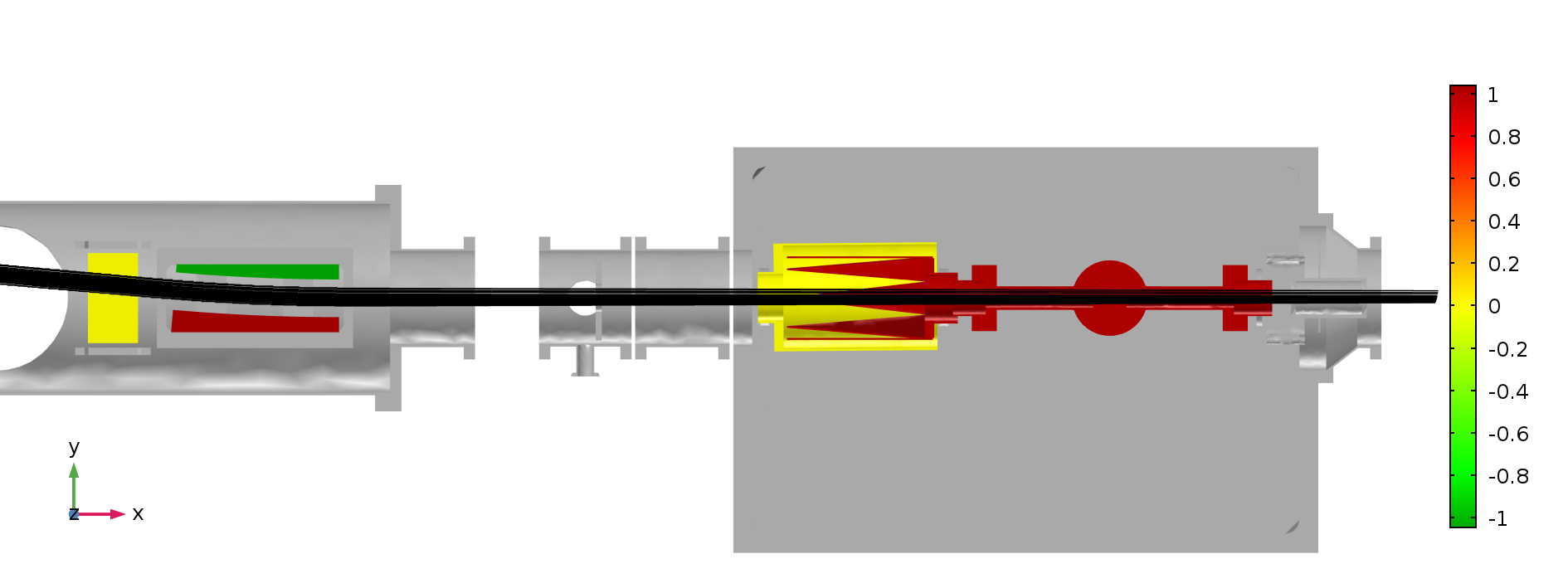}};
        \draw (7.2, -0.6) node[rotate=-90, text width=5cm] {Electric potential [kV]};
\end{tikzpicture}
\caption{Simplified beamline geometry as used in the COMSOL simulations. The electrodes are colored with the applied electric potential when the scanner is set to 1 kV (color online). The other (grounded) beamline elements present in the simulation are colored gray. The trajectories of the particles are drawn in black.}
\label{fig:comsolgeom}
\end{figure*}
In the interaction region, where the optical pumping takes place, Helmholtz coils provide a magnetic field on the order of 2 mT along the beamline axis pointing in the beam direction. This magnetic field defines a quantisation axis and avoids coupling of the atomic spins to possible stray fields in the environment. The minimal length of this interaction region is determined by the time needed for the pumping process. The process of optical pumping can be modelled through the formation of rate equations based on the Einstein formalism \cite{Foot2005a,Ruiz2017,Yordanov2007}. By solving this system of differential equations, the degree of nuclear polarization $P$ (Eq.~\eqref{eq:pol}) can be calculated for any interaction time. The D2 line in Na \cite{0954-3899-44-8-084005} was used as the case study. Figure~\ref{fig:designcalc}(a) shows a the hyperfine spectrum generated with this method, while in (b) the polarization in the highest peak is calculated as a function of the laser-atom interaction time (translated into a beam line length assuming a 50 keV beam). Based on these calculations, a length of 1.6 m was selected as a compromise between achievable polarization and available space in the ISOLDE hall. Although the length needed to fully polarize an ensemble depends on the Einstein $A$ parameter of the transition, 1.6 m will give a sufficiently long interaction time for most strong transitions where $A$ has a value in the order of $10^7-10^8$ Hz. This is illustrated in Figure~\ref{fig:designcalc}(b) for $^{26,28}$Na and $^{35}$Ar, the first isotopes that have been polarized with the new set-up.

A series of solenoids and a large electromagnet are placed after the interaction region, with the field of the solenoids acting along the beam direction and the electromagnet generating a field perpendicular to it. The combination of these fields adiabatically rotates the atomic spin in the horizontal plane, orienting it in the same direction as the field of the electromagnet. The field of the electromagnet is sufficiently strong to decouple the nuclear from the electron spin. The details of this adiabatic spin rotation and decoupling of the nuclear spin from the electron spin are discussed in Section~\ref{sec:adiabaticrotation}. A removable sample holder and \textbeta-detectors are installed inside the electromagnet, where the polarized ensemble is implanted in a suitable host material. This \textbeta-detection setup has been used before in \textbeta-NMR studies on Mg isotopes \cite{Kowalska2008}.

A diagnostics box containing a wire scanner and Faraday cup, as detailed in \cite{Kugler2000} (supplemented with a plate to detect atomic beams), is installed after this region to provide beam diagnostics.

\section{Ion optics}\label{sec:beamsim}

\begin{figure}[!b]
\centering
    \includegraphics[width=0.75\columnwidth]{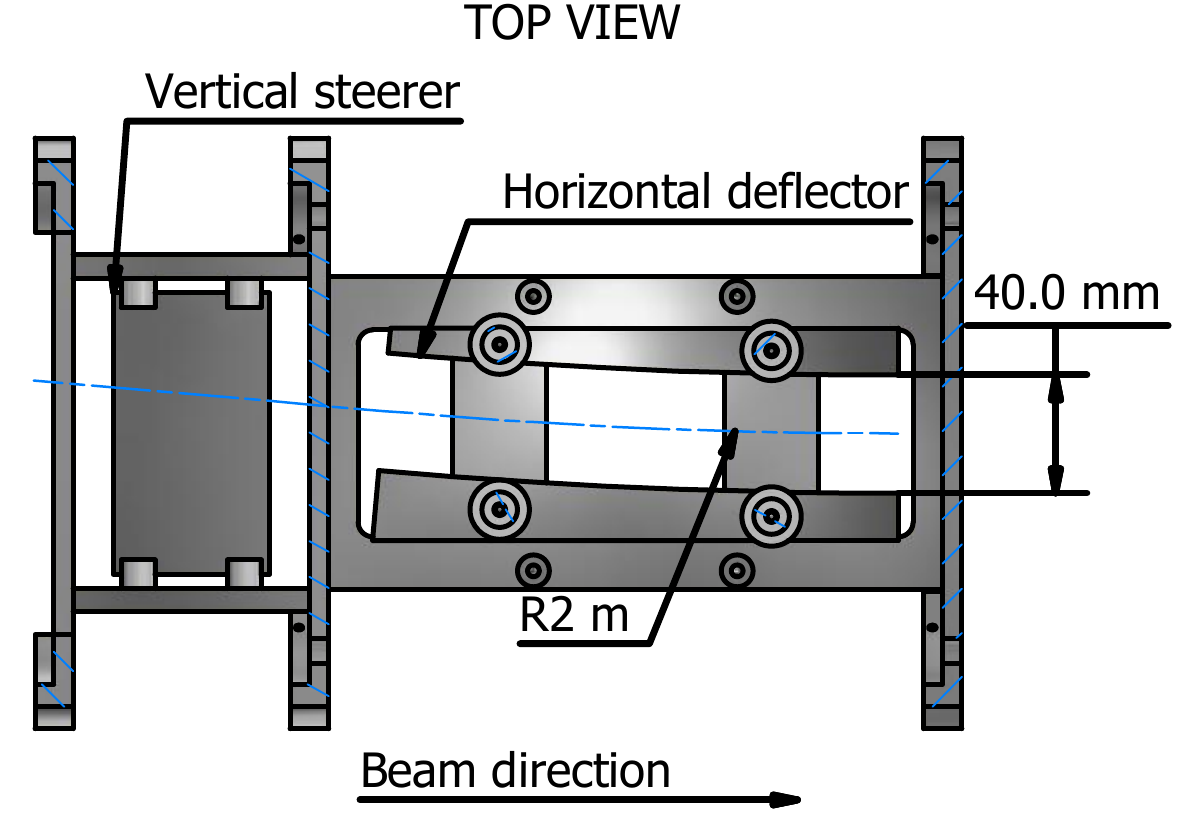}
    \caption{CAD drawing of the 5 degree ion-deflector. Ion beam coming from ISOLDE (left) is bent 5 degrees with a radius of 2 m to overlap the ion/atom beam with the laser beam.}
    \label{fig:deflectordesign}
\end{figure}

\begin{figure*}[!ht]
\centering
\subfloat[]{
\includegraphics[width=0.75\textwidth]{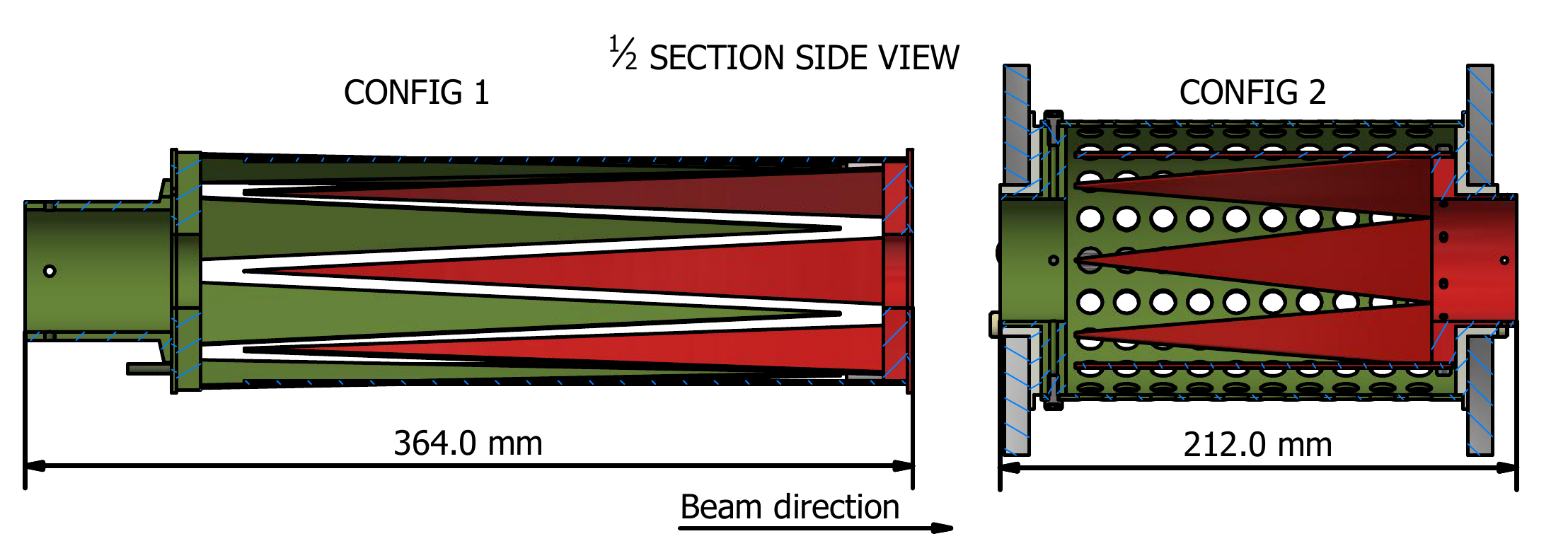}\label{fig:voltscanners}
}\\
\subfloat[]{\includegraphics[width=0.40\textwidth]{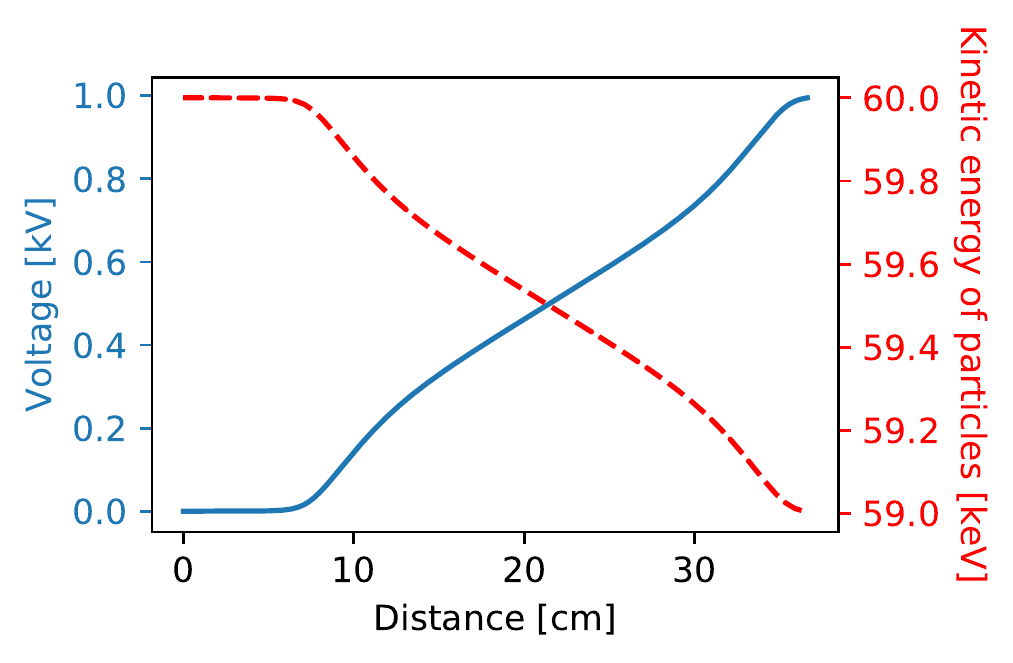}\label{fig:angle_old}}
\subfloat[]{\includegraphics[width=0.40\textwidth]{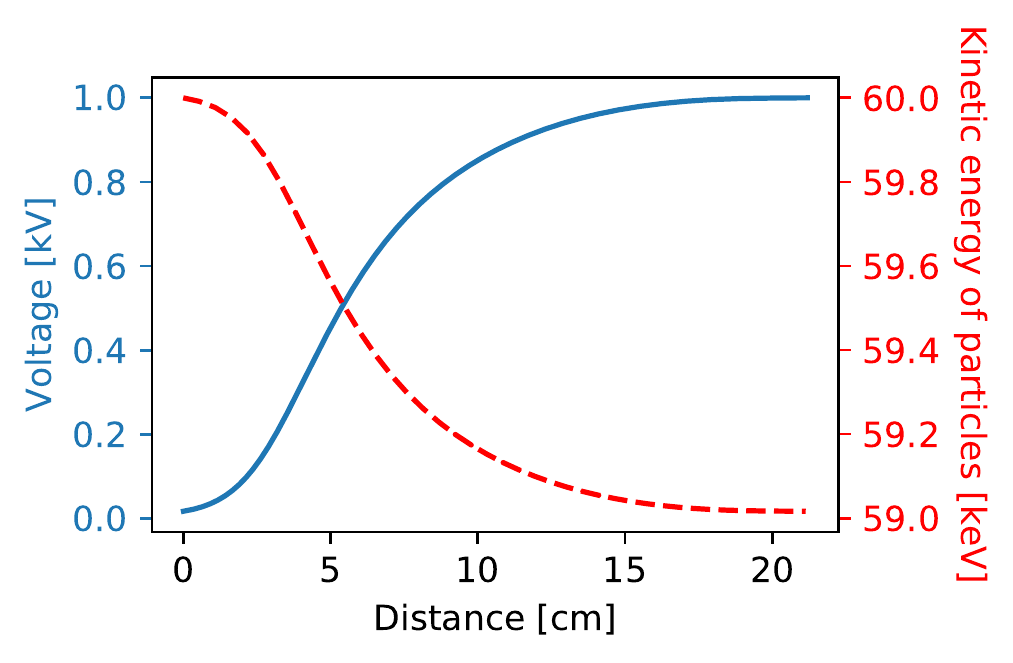}\label{fig:angle_new}}
\caption{(a) 1/2 section view of both configurations for the voltage scanner. The biased electrode is indicated in red and the grounded electrode in green (color online). (b) Electric potential (full line) and the kinetic energy (dashed) of a 60 keV beam of $^{39}$K as a function of distance using configuration 1 when 1 kV is applied to the biased electrode. (c) Same as (b), but for configuration 2.}
\label{fig:voltscan}
\end{figure*}

Ion-optical beam transmission simulations were performed to benchmark the effect that the deflector and voltage scanner have on the shape of the beam and the transmission that can be expected.

A standard beam of $^{39}$K$^+$ with 3 \textpi{} mm mrad beam emittance is generated for the simulations, as reported for the cooler-buncher ISCOOL \cite{Mane2009}. The focal point of the beam is, prior to entering the 5 degree deflector, optimized for maximal transmission using the quadrupole doublet that is installed in front of it. As the doublet is not included in the simulations, the focus is set by changing the size of the beam and the convergence of the incoming beam by changing the Twiss parameters \cite{Floetmann2003}. The kinetic energy of the beam is set to 60~keV, which is the maximal beam energy that can be delivered by ISOLDE to the low energy section. For all simulations, the COMSOL multiphysics software was used \cite{comsol}. The used geometry is given in Figure~\ref{fig:comsolgeom}. First an overview of all the elements included in the simulations will be given.

The first electrostatic element of the VITO beamline in the simulations is the 5 degree deflector, which allows a soft bending of the ion beam to overlap it collinearly with the laser light. The 5 degree deflector has an internal opening of 40 mm and consists of two vertical steerer plates and a pair of electrodes with a machined curve matching 2 m (see Figure~\ref{fig:deflectordesign}). After this, a voltage scanner (the design of which will be treated further in this section) adjusts the kinetic energy of the ion beam in a range of $\pm 1$ keV. It is mounted inside the vacuum box where the CEC is also mounted (shaped electrodes and red circle respectively, Figure~\ref{fig:comsolgeom}) and the biased electrode is connected to the CEC. The CEC acts as a long collimator with a 2 cm opening, followed by another collimator with an opening of 1 cm approximately 2 m further downstream. These small collimators guarantee a good overlap between the particle and laser beam. As the charge exchange process neutralizes the charged particles, the CEC is the final electrostatic element considered in the simulations. The tubes and chambers forming the beamline up to this point are also present and are grounded to provide accurate potential fields.

The design for the voltage scanner deviates from a traditional series of ring electrodes connected through a resistor chain \cite{Ruiz2017}. Instead, two specially shaped electrodes define the equipotential electrical field (Figure~\ref{fig:voltscanners}). Two configurations of this geometrical design of the shaped electrodes have been used.

In the first configuration (Config. 1 in Figure~\ref{fig:voltscanners}), eight triangular spikes are attached to an octagonal mounting base. Overlapping two of such electrodes gives a gradual and nearly-linear change in potential experienced by the ions, as given in Figure~\ref{fig:angle_old}. In the second design (Config. 2 in Figure~\ref{fig:voltscanners}), the first electrode is replaced with a cylinder covering the entire scanner, separated from the biased electrode with a teflon insulator. The central beam axis is thus better encapsulated, resulting in a more sigmoidal variation of the electric potential (Figure~\ref{fig:angle_new}). To emulate mechanical imperfections related to the construction, the grounding electrode was rotated 0.5$^\circ$ relative to the z-axis as defined in Figure~\ref{fig:comsolgeom}. This rotation is present in all simulations for both configurations.

\begin{figure}[!t]
\centering
\includegraphics[width=\columnwidth]{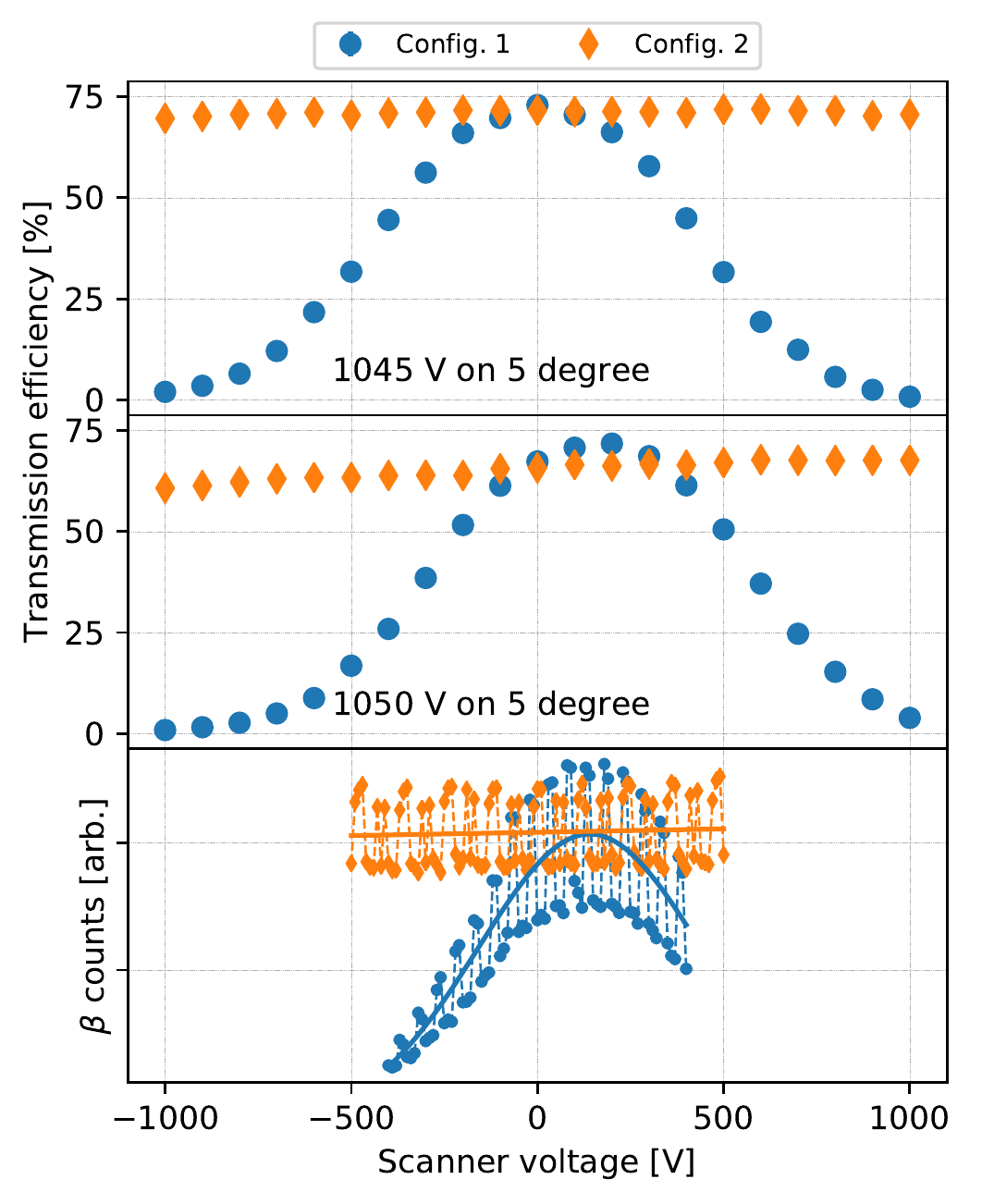}\label{fig:transmissioncomparisonstraight}
\caption{Transmission efficiency through the beamline, simulated for both configurations of the scanning electrodes, as a function of voltage applied to the scanner and for two different voltages on the deflector. When the optimal voltage is applied to the 5 degree deflector (1045 V, top plot), the beam is centered on the axis of the biased electrode, while a slightly offset voltage (1050 V, middle plot) results in slightly lower transmission since the axes of the beam and the electrode no longer coincide. The experimental transmission (bottom plot) agrees with the general trend predicted by the simulations. The staggering in the data is due to the structure of the proton supercycle.}
\label{fig:transmissioncomparison}
\end{figure}

The transmission of the beam through the beam line as a function of scanning voltage has been simulated for the two different designs of the voltage scanner (Figure~\ref{fig:transmissioncomparison}a). A second set of simulations were performed with a slightly detuned 5 degree deflector (Figure~\ref{fig:transmissioncomparison}b). Both designs have been constructed and used in experiments, such that the transmission simulations can be compared to actual data (Figure~\ref{fig:transmissioncomparison}c).

The transmission data for the first configuration was gathered in the commissioning experiment \cite{0954-3899-44-8-084005} and were the total \textbeta-counts from a $^{26}$Na beam measured as a function of scanning voltage. During an experiment on \textbeta-NMR in liquid samples performed in May 2018 \cite{Kowalska:2299798}, the second configuration was used for the first time and the data was gathered in the same way.

The simulations (top 2 panels of Figure~\ref{fig:transmissioncomparison}) and the online data (bottom panel of Figure~\ref{fig:transmissioncomparison}) show a very good agreement. The oscillation in the counts is due to the proton supercycle. In both the measurements and the simulations, the first configuration has a severe beamsteering effect, reducing the transmission efficiency as a function of scanning voltage. The second configuration has no such effect, owing to the better encapsulation of the beam axis. The simulations with a detuned 5 degree deflector show that this can result in a slight slope for the second configuration and a shift in the peak location for the first. These features are also present in the data, suggesting an almost perfect beamtune during the experiment.

\section{Design of magnetic fields}\label{sec:fields}

\begin{table*}[!ht]
    \caption[Optimized coil parameters.]{Optimized parameters for the coils as determined with COMSOL simulations. A wire thickness of 0.8 mm was used through all simulations. For the Helmholtz coils, the radius refers to the inscribed circle. The length refers to the dimension of the coil along the beamline axis. The number after the description refers to the labelling in Figure~\ref{fig:beamline}.}
    \label{tab:solenoids}
    \centering
    \begin{tabular}{lrrrrrrr}
    \toprule
& Helmholtz coils (1) & Large solenoid (2) & \multicolumn{5}{c}{Small solenoids (3)}\\
&  &  & 1 &  2 &  3 &  4 &  5\\
    \midrule
        Current [A] & 1.6 & 1.0 & \multicolumn{5}{c}{2} \\
        Windings [\#] & 1000 & 900 & 10&65&154&1100&225 \\
        \addlinespace
        Radius [mm] & 400 & 112.5 & \multicolumn{5}{c}{20.5}\\
        Length [mm] & 32 & 315 & 40 & 60 & 40 & 154 & 66 \\
    \addlinespace
    \bottomrule
    \end{tabular}
\end{table*}

The magnetic field generated in the beamline is separated in three separate sections (see Figure~\ref{fig:beamline}), each with their own requirements:
\begin{enumerate}
    \item the interaction region (labelled 1 and 2 in Figure~\ref{fig:beamline}).
    \item the transitional field region (labelled 3).
    \item the main magnet region.
\end{enumerate}
The interaction region has to provide a weak, uniform magnetic field over the entire beam path that needs to compensate for stray fields in order to maintain the laser-induced atomic spin polarization. The field should be small enough however, not to induce a large splitting of the magnetic substates of the hyperfine levels. A magnetic field of approximately 2 mT fulfills both requirements.

Once the radioactive beam is implanted, the magnetic field has to be strong enough to decouple the nuclear spin from random interactions with potential (defect-associated) electric field gradients in the crystal. The installed electromagnet can generate a field of up to 0.7 T. This value depends on both the current supplied to the magnet as well as the distance between the magnet poles, which can be varied. With a maximal pole distance of 8 cm, different setups for holding samples and placing detectors can be accommodated.

Since the field generated by the electromagnet is perpendicular to the beamline axis (which is also the atomic spin orientation axis), the transitional field region has to be tuned to provide adiabatic rotation of the oriented atomic spins. The field previously designed for the \textbeta-NMR setup at COLLAPS (see Ref. \cite{matthias1996}) was used as the model field.

For the design of the magnetic field of both the interaction and the transitional region, simulations were made in COMSOL. As a design choice, 4 octagonal coils arranged in a Helmholtz configuration are used for the interaction region (see Figure~\ref{fig:beamline}). Due to geometrical concerns, a solenoid with 11.25 cm radius extends the interaction region. Several smaller solenoids directly wound onto a beampipe continue past this point and form the transitional field. The final parameters of all solenoids are given in Table~\ref{tab:solenoids}.

\begin{figure}[t!]
\centering
    \includegraphics[width=\columnwidth]{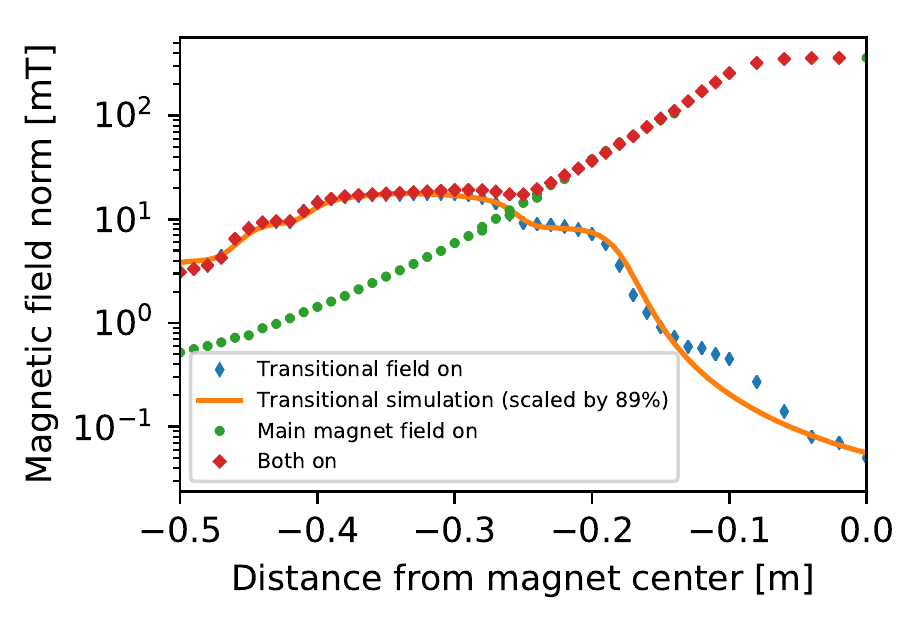}
    \caption{Comparison between the simulation (solid line) and measurements (markers) of the magnetic field. The simulation is scaled down by a global factor to account for a difference between the read out current and the current applied to the coils. The discrepancy above -0.2 m is due to the residual magnetization of the magnet.}
    \label{fig:magneticfields}
\end{figure}

In order to compare the simulated magnetic field profile with reality, magnetic field measurements with a 3D Hall probe were made in three circumstances: the transitional field and the main magnet on (red wide diamonds in Figure~\ref{fig:magneticfields}), only the main magnet powered (green dots) and only the transitional field powered (blue thin diamonds). Due to the unknown configuration in the main magnet, only the transitional field could be simulated in COMSOL (full line). The good agreement between simulation and measurements suggests a good correspondence between the coil parameters in the simulation and the physical coils.

\section{Multiple-frequency pumping}\label{sec:multipump}

This can be overcome by using multi-frequency optical pumping to simultaneously excite more than one hyperfine transitions. The same concept has already been applied at the TRIUMF facility for the polarization of Li, where EOM's were used to induce side-frequencies in the range of $\pm 30$ MHz to the main laser beam frequency \cite{Levy2010}.

\begin{table}[t!]
    \caption{The hyperfine parameters of $^{35}$Ar, deduced from the measured hyperfine parameters of $^{39}$Ar in the same laser transition \cite{Weite2010} and the known $^{35}$Ar nuclear moments \cite{2016_Mertzimekis}.}
    \label{tab:hyperfineparams}
    \centering

    \resizebox{\columnwidth}{!}{
        \begin{tabular}{lcc}
        \toprule
        & A [MHz] (calc.) & B [MHz] (calc.)\\
        \midrule
        $1s_5$ ($J=2$) & 265.8(28) & 83(25) \\
        $2p_9$ ($J=3$) & 125.6(12) & 80(19) \\
        \bottomrule
        \end{tabular}
        }
\end{table}

We have verified this experimentally, using a $^{35}$Ar beam produced by a 1.4 GeV beam onto a CaO target at ISOLDE. The multi-frequency pumping was realized by using two acousto-optic modulators (AOMs), which can each produce a side band frequency in the required ranges of 300-400 MHz by applying a fixed RF-frequency to the crystal inside the AOM. This shifts the frequency of the incoming laser light by the same value. For details on the setup of the AOMs, see Ref.~\cite{Gins2018b}.

\begin{figure}[t!]
    \centering
    \includegraphics[width=\columnwidth]{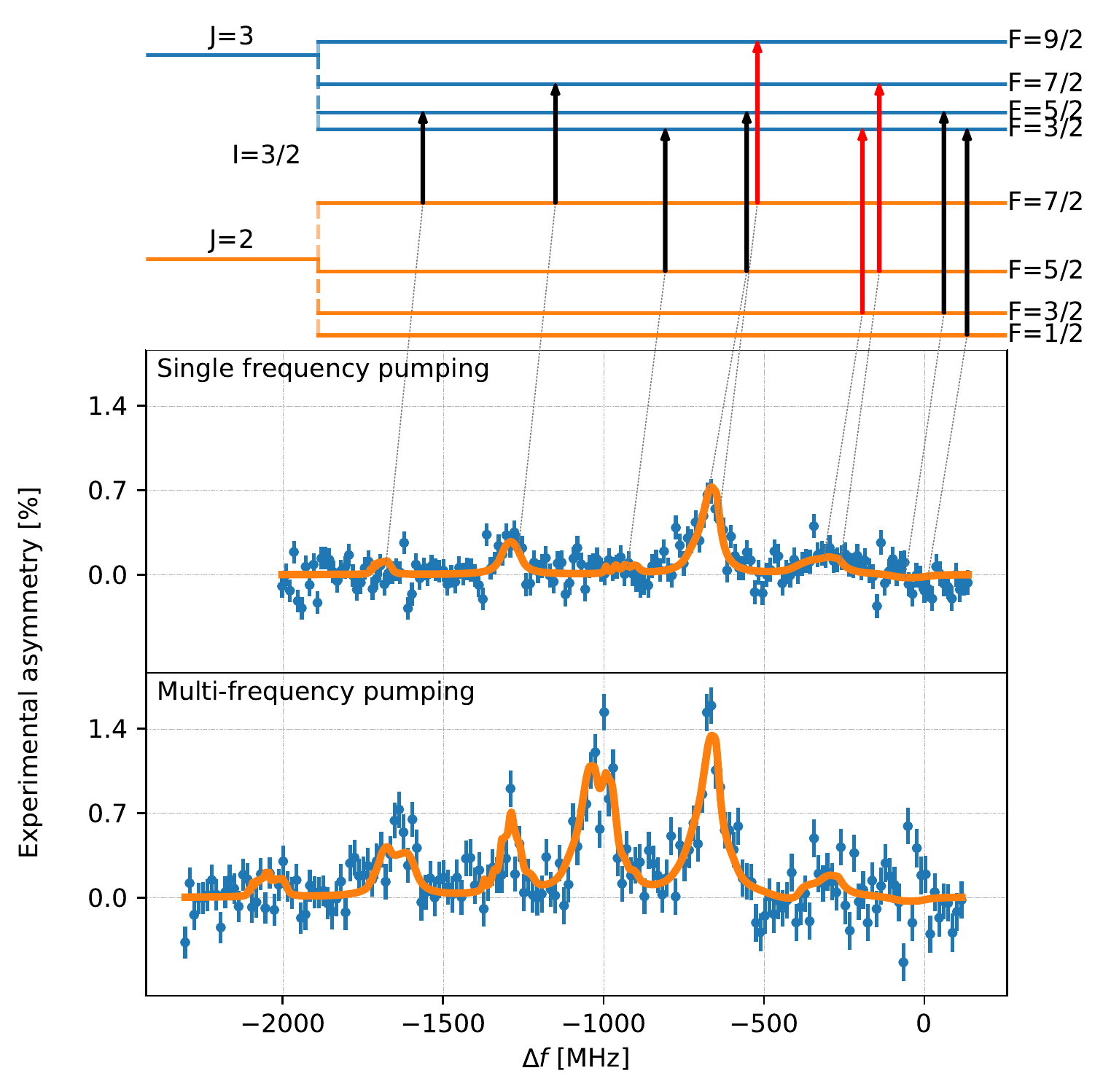}
    \caption{Top: Hyperfine structure of the 811 nm transition with indicated allowed transitions. Using the hyperfine transitions $7/2\rightarrow 9/2$, $5/2\rightarrow 7/2$ and $3/2\rightarrow 3/2$ (marked in red) saturates the polarization of the ensemble. Bottom plots: experimental spectra for both single- and multi-frequency pumping. Both spectra were fitted simultaneously to the rate equations. The laser power in the main beam was set to be a shared fit parameter, while the hyperfine parameters were fixed to the derived value (see Table~\ref{tab:hyperfineparams}). The transitions in the top are connected with their location in the single-frequency pumping spectrum.}
    \label{fig:multipump}
\end{figure}

The observed hyperfine spectra, using either one laser frequency for the scanning or by scanning the 3 laser frequencies simultaneously, are shown in Figure~\ref{fig:multipump}. The induced nuclear spin-polarization is observed by measuring the asymmetry in the radioactive \textbeta-decay after implantation in a suitable crystal, as outlined before. The spectra shown in Figure~\ref{fig:multipump} have been recorded by implanting the polarized $^{35}$Ar beam into a NaCl crystal kept at a temperature of 15(5) K. The signal height in both spectra is significantly different, as can be seen clearly in e.g. the $7/2\rightarrow 9/2$ transition, for which the observed asymmetry increases by almost a factor of 2.

In order to quantify the observed gain in signal strength (and thus spin-polarization), the data have been fitted using the rate equations that were implemented using the SATLAS Python package \cite{Gins2018}. In the fitting procedure, the laser power (determining the linewidths), the centroid of the spectrum and the total signal height (proportional to the induced nuclear polarization) were left as free parameters, while the A and B factors as given in Table~\ref{tab:hyperfineparams} were kept as fixed values. Both spectra were fit simultaneously with the same value for the laser power, thus ensuring the correlations due to shared parameters were propagated correctly when determining the ratio of the signal strength.

An excellent agreement is found between the observed spectra and the calculated \textbeta-asymmetry spectra as a function of the laser frequency, both for the single and triple laser-atom interaction systems. This gives confidence in the predictive power of this simulation package, such that it can be used in the future to optimize the laser polarization experiments. From the fitted signal strengths in both spectra, we find an increase of a factor 1.85(3)

\section{Adiabatic rotation}
\label{sec:adiabaticrotation}

In the experimental set-up, the atomic spin-polarization symmetry axis is along the (laser) beam line. On the other hand, for the \textbeta-asymmetry measurements, the nuclear spin-polarization symmetry axis should be along the direction of the strong holding field in which the implantation crystal is mounted. This field is perpendicular to the beam line, in order to allow for \textbeta-detectors to be mounted at 0 and 180 degrees with respect to the field direction. From the rate equation calculations of the atomic population, the nuclear spin polarization is extracted under the assumption that the decoupling field is oriented along the symmetry axis. As we apply a gradually increasing magnetic field in order to rotate (and decouple) the nuclear and electron spins adiabatically into the main field direction, changes in the nuclear spin polarization due to the adiabtic rotation process are possible. To this end, simulations of this adiabatic spin rotation have been performed. The magnetic field profile in the 3 directions has been measured and used in these simulations, for which the total field strength along the beam line was shown in Figure~\ref{fig:magneticfields}.

Quantum mechanical calculations, starting from the interaction Hamiltonian including the hyperfine interaction and the two Zeemann interactions with $\vec{I}$ and $\vec{J}$, were performed with QuTiP \cite{johansson2013}. The state vector is initialized in the atomic groundstate with populations in the $\vec{F}$ and $m_F$ states as given by the rate equations after the optical pumping process. The interaction strength of the magnetic field in the three directions is made time-dependent, representing the changing magnetic field as the particle beam travels through the setup at a certain speed. The spin dynamics are then calculated by solving the Schr\"odinger equation. Experimental values for nuclear parameters in these equations were taken from Refs.~\cite{Touchard1982} and \cite{2016_Mertzimekis}. The calculated flight time for the beam from the start of the transitional magnetic field to the implantation host is extended to also include a period where the beam is stopped in the host. The first period (the \emph{dynamic} field region) is 2/3 of the total time solved for, while the implanted period (the \emph{static} field region) accounts for 1/3 of the time. More details of the calculation can be found in Ref.~\cite{Gins2018b}.

\begin{figure}[t]
\centering
    \includegraphics[width=\columnwidth]{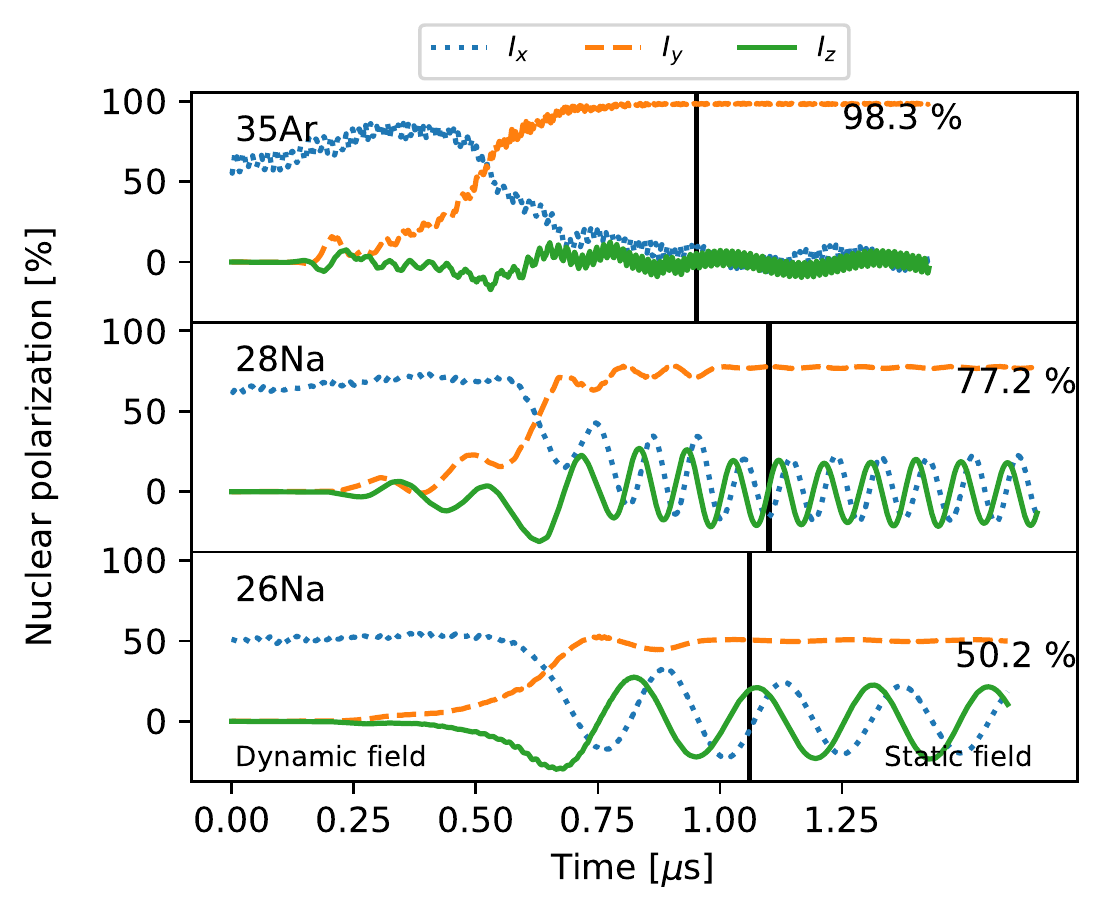}
    \caption{Calculated change in the nuclear spin polarization along the $x$, $y$ and $z$ directions (dotted and dashed and full lines) due to the spin rotation from along the beam axis (x) to along the main field axis (y).}
    \label{fig:rotationsim}
\end{figure}

Table~\ref{tab:rotationvalues} presents the calculated nuclear spin polarization from both the rate equations and the quantum mechanical calculations. The observed asymmetry ratio between $^{26}$Na and $^{28}$Na matches the quantum mechanical calculation, although the absolute number is too high.

Simulations have also been made for the adiabatic rotation of $^{35}$Ar (top panel of Figure~\ref{fig:rotationsim}), which shows no loss in nuclear spin polarization from the rotation process.

\begin{table}[t]
    \caption{Calculated and observed nuclear spin polarization of $^{26,28}$Na. Experimental data taken from Ref.~\cite{0954-3899-44-8-084005}.}
    \label{tab:rotationvalues}
    \centering
\resizebox{\columnwidth}{!}{
    \begin{tabular}{lccc}
    \toprule
    & QM calc. & Rate equation & Experiment\\
    \midrule
    \addlinespace
    \midrule
    \addlinespace
    Ratio & 1.54 & 1.43 & 1.51\\
    \bottomrule
    \end{tabular}
    }
\end{table}

\section{Conclusion}\label{sec:conclusion}

To answer the demand for accessible spin-polarized radioactive nuclei, the VITO beamline was built as a dedicated setup. It is an adaptable beamline that delivers highly polarized nuclei to a central detection point.

Multi-frequency pumping has been established as a viable technique to increase the asymmetry signal that can be expected from the radioactive species. Tests on $^{35}$Ar show that the increase and spectrum can be reproduced by the rate equation model.

The series of magnetic fields provide an efficient adiabatic rotation and decoupling of the nuclear spin. Rotation calculations agree with the observed asymmetry ratios. The rotation calculations can be repeated for different species provided the necessary nuclear parameters are available.

\section{Acknowledgments}

This work was supported by the ERC Starting Grant no. 640465, FWO-Vlaanderen (Belgium) and GOA 15/010 from KU Leuven, the Science and Technology Facilities Council (UK).
\bibliographystyle{model1a-num-names}
\bibliography{vito_technics}
\end{document}

%% file: pumping.txt
\begin{tikzpicture}[
upperlevel/.style={black,very thick},
lowerlevel/.style={black,very thick},
splitting/.style={densely dashed},
label/.style={black,font=\tiny\bf},
excitation/.style={->,>=stealth,very thick},
decay/.style={->, >=stealth,densely dashed},
scale=1,
]

\pgfmathsetmacro{\levellength}{0.7};
\pgfmathsetmacro{\splittingmultiplier}{1.5};
\pgfmathsetmacro{\levelseperation}{0.25};
\pgfmathsetmacro{\magneticseperation}{0.1};
\pgfmathsetmacro{\magnetstart}{1.15}

\pgfmathsetmacro{\lowerstart}{0};
\pgfmathsetmacro{\upperstart}{1};

\pgfmathsetmacro{\lowerone}{-0.26666*\splittingmultiplier};
\pgfmathsetmacro{\lowertwo}{0.13334*\splittingmultiplier};

\pgfmathsetmacro{\upperone}{-0.24432296*\splittingmultiplier};
\pgfmathsetmacro{\uppertwo}{-0.11135416*\splittingmultiplier};
\pgfmathsetmacro{\upperthree}{0.15567704*\splittingmultiplier};

\draw[lowerlevel] (0,\lowerstart) node[left,label]{J=1/2}-- (\levellength,\lowerstart);
\draw[upperlevel] (0,\upperstart) node[left,label]{J=3/2}-- (\levellength,\upperstart);

\node[label] () at (0,0.5*\lowerstart+0.5*\upperstart) {I=1};

\draw[lowerlevel] (\levellength+\levelseperation,\lowerstart+\lowerone) -- +(\levellength,0) node[right,label]{F=1/2};
\draw[splitting] (\levellength,\lowerstart) -- +(\levelseperation,\lowerone);
\draw[lowerlevel] (\levellength+\levelseperation,\lowerstart+\lowertwo) -- +(\levellength,0) node[right,label]{F=3/2};
\draw[splitting] (\levellength,\lowerstart) -- +(\levelseperation,\lowertwo);

\draw[lowerlevel] (3*\levellength+\levelseperation+\magnetstart+1*\magneticseperation,\lowerstart+\lowertwo) -- +(\levellength,0) node[midway,below,label,align=center]{$m_F$\\-3/2};
\draw[lowerlevel] (4*\levellength+\levelseperation+\magnetstart+2*\magneticseperation,\lowerstart+\lowertwo) -- +(\levellength,0) node[midway,below,label,align=center]{$m_F$\\-1/2};
\draw[lowerlevel] (5*\levellength+\levelseperation+\magnetstart+3*\magneticseperation,\lowerstart+\lowertwo) -- +(\levellength,0) node[midway,below,label,align=center]{$m_F$\\1/2};
\draw[lowerlevel] (6*\levellength+\levelseperation+\magnetstart+4*\magneticseperation,\lowerstart+\lowertwo) -- +(\levellength,0) node[midway,below,label,align=center]{$m_F$\\3/2};

\draw[upperlevel] (\levellength+\levelseperation,\upperstart+\upperone) -- +(\levellength,0) node[right,label]{F=1/2};
\draw[splitting] (\levellength,\upperstart) -- +(\levelseperation,\upperone);
\draw[upperlevel] (\levellength+\levelseperation,\upperstart+\uppertwo) -- +(\levellength,0) node[right,label]{F=3/2};
\draw[splitting] (\levellength,\upperstart) -- +(\levelseperation,\uppertwo);
\draw[upperlevel] (\levellength+\levelseperation,\upperstart+\upperthree) -- +(\levellength,0) node[right,label]{F=5/2};
\draw[splitting] (\levellength,\upperstart) -- +(\levelseperation,\upperthree);

\draw[upperlevel] (2*\levellength+\levelseperation+\magnetstart+0*\magneticseperation,\upperstart+\upperthree) -- +(\levellength,0) node[midway,above,label,align=center]{$m_F$\\-5/2};
\draw[upperlevel] (3*\levellength+\levelseperation+\magnetstart+1*\magneticseperation,\upperstart+\upperthree) -- +(\levellength,0) node[midway,above,label,align=center]{$m_F$\\-3/2};
\draw[upperlevel] (4*\levellength+\levelseperation+\magnetstart+2*\magneticseperation,\upperstart+\upperthree) -- +(\levellength,0) node[midway,above,label,align=center]{$m_F$\\-1/2};
\draw[upperlevel] (5*\levellength+\levelseperation+\magnetstart+3*\magneticseperation,\upperstart+\upperthree) -- +(\levellength,0) node[midway,above,label,align=center]{$m_F$\\1/2};
\draw[upperlevel] (6*\levellength+\levelseperation+\magnetstart+4*\magneticseperation,\upperstart+\upperthree) -- +(\levellength,0) node[midway,above,label,align=center]{$m_F$\\3/2};
\draw[upperlevel] (7*\levellength+\levelseperation+\magnetstart+5*\magneticseperation,\upperstart+\upperthree) -- +(\levellength,0) node[midway,above,label,align=center]{$m_F$\\5/2};

\draw[excitation] (3.5*\levellength+\levelseperation+\magnetstart+1*\magneticseperation,\lowerstart+\lowertwo) -- +(\levellength+\magneticseperation,\upperstart+\upperthree-\lowerstart-\lowertwo);
\draw[excitation] (4.5*\levellength+\levelseperation+\magnetstart+2*\magneticseperation,\lowerstart+\lowertwo) -- +(\levellength+\magneticseperation,\upperstart+\upperthree-\lowerstart-\lowertwo);
\draw[excitation] (5.5*\levellength+\levelseperation+\magnetstart+3*\magneticseperation,\lowerstart+\lowertwo) -- +(\levellength+\magneticseperation,\upperstart+\upperthree-\lowerstart-\lowertwo);
\draw[excitation] (6.5*\levellength+\levelseperation+\magnetstart+4*\magneticseperation,\lowerstart+\lowertwo) -- +(\levellength+\magneticseperation,\upperstart+\upperthree-\lowerstart-\lowertwo);

\draw[decay] (2.5*\levellength+\levelseperation+\magnetstart+0*\magneticseperation,\upperstart+\upperthree) -- +(\levellength,-\upperstart-\upperthree+\lowerstart+\lowertwo);

\draw[decay] (3.5*\levellength+\levelseperation+\magnetstart+1*\magneticseperation,\upperstart+\upperthree) -- +(0,-\upperstart-\upperthree+\lowerstart+\lowertwo);
\draw[decay] (3.5*\levellength+\levelseperation+\magnetstart+1*\magneticseperation,\upperstart+\upperthree) -- +(\levellength,-\upperstart-\upperthree+\lowerstart+\lowertwo);

\draw[decay] (4.5*\levellength+\levelseperation+\magnetstart+2*\magneticseperation,\upperstart+\upperthree) -- +(-\levellength,-\upperstart-\upperthree+\lowerstart+\lowertwo);
\draw[decay] (4.5*\levellength+\levelseperation+\magnetstart+2*\magneticseperation,\upperstart+\upperthree) -- +(0,-\upperstart-\upperthree+\lowerstart+\lowertwo);
\draw[decay] (4.5*\levellength+\levelseperation+\magnetstart+2*\magneticseperation,\upperstart+\upperthree) -- +(\levellength,-\upperstart-\upperthree+\lowerstart+\lowertwo);

\draw[decay] (5.5*\levellength+\levelseperation+\magnetstart+3*\magneticseperation,\upperstart+\upperthree) -- +(-\levellength,-\upperstart-\upperthree+\lowerstart+\lowertwo);
\draw[decay] (5.5*\levellength+\levelseperation+\magnetstart+3*\magneticseperation,\upperstart+\upperthree) -- +(0,-\upperstart-\upperthree+\lowerstart+\lowertwo);
\draw[decay] (5.5*\levellength+\levelseperation+\magnetstart+3*\magneticseperation,\upperstart+\upperthree) -- +(\levellength,-\upperstart-\upperthree+\lowerstart+\lowertwo);

\draw[decay] (6.5*\levellength+\levelseperation+\magnetstart+4*\magneticseperation,\upperstart+\upperthree) -- +(-\levellength,-\upperstart-\upperthree+\lowerstart+\lowertwo);
\draw[decay] (6.5*\levellength+\levelseperation+\magnetstart+4*\magneticseperation,\upperstart+\upperthree) -- +(0,-\upperstart-\upperthree+\lowerstart+\lowertwo);

\draw[decay] (7.5*\levellength+\levelseperation+\magnetstart+5*\magneticseperation,\upperstart+\upperthree) -- +(-\levellength,-\upperstart-\upperthree+\lowerstart+\lowertwo);

\node[label] at (2.5*\levellength+\levelseperation+\magnetstart+0*\magneticseperation,0.5*\upperstart+0.5*\upperthree+0.5*\lowerstart+0.5*\lowertwo) {$\sigma^+$};

\draw[excitation] (1.05*\levellength+\levelseperation, \lowerstart+\lowerone) -- (1.05*\levellength+\levelseperation, \upperstart+\upperone);
\draw[excitation] (1.25*\levellength+\levelseperation, \lowerstart+\lowerone) -- (1.25*\levellength+\levelseperation, \upperstart+\uppertwo);

\draw[excitation] (1.55*\levellength+\levelseperation, \lowerstart+\lowertwo) -- (1.55*\levellength+\levelseperation, \upperstart+\upperone);
\draw[excitation] (1.75*\levellength+\levelseperation, \lowerstart+\lowertwo) -- (1.75*\levellength+\levelseperation, \upperstart+\uppertwo);
\draw[excitation] (1.95*\levellength+\levelseperation, \lowerstart+\lowertwo) -- (1.95*\levellength+\levelseperation, \upperstart+\upperthree);

\end{tikzpicture}

%% file: main.bbl
\begin{thebibliography}{27}
\expandafter\ifx\csname natexlab\endcsname\relax\def\natexlab#1{#1}\fi
\providecommand{\url}[1]{\texttt{#1}}
\providecommand{\href}[2]{#2}
\providecommand{\path}[1]{#1}
\providecommand{\DOIprefix}{doi:}
\providecommand{\ArXivprefix}{arXiv:}
\providecommand{\URLprefix}{URL: }
\providecommand{\Pubmedprefix}{pmid:}
\providecommand{\doi}[1]{\href{http://dx.doi.org/#1}{\path{#1}}}
\providecommand{\Pubmed}[1]{\href{pmid:#1}{\path{#1}}}
\providecommand{\bibinfo}[2]{#2}
\ifx\xfnm\relax \def\xfnm[#1]{\unskip,\space#1}\fi
\bibitem[{Kowalska et~al.(2017)Kowalska, Aschenbrenner, Baranowski, Bissell,
  Gins, Harding, Heylen, Neyens, Pallada, Severijns, Velten, Walczak,
  Wienholtz, Xu, Yang, and Zakoucky}]{0954-3899-44-8-084005}
\bibinfo{author}{M.~Kowalska}, \bibinfo{author}{P.~Aschenbrenner},
  \bibinfo{author}{M.~Baranowski}, \bibinfo{author}{M.~L. Bissell},
  \bibinfo{author}{W.~Gins}, \bibinfo{author}{R.~D. Harding},
  \bibinfo{author}{H.~Heylen}, \bibinfo{author}{G.~Neyens},
  \bibinfo{author}{S.~Pallada}, \bibinfo{author}{N.~Severijns},
  \bibinfo{author}{P.~Velten}, \bibinfo{author}{M.~Walczak},
  \bibinfo{author}{F.~Wienholtz}, \bibinfo{author}{Z.~Y. Xu},
  \bibinfo{author}{X.~F. Yang}, \bibinfo{author}{D.~Zakoucky},
  \bibinfo{journal}{Journal of Physics G: Nuclear and Particle Physics}
  \bibinfo{volume}{44} (\bibinfo{year}{2017}) \bibinfo{pages}{084005}.
\bibitem[{Wu et~al.(1957)Wu, Ambler, Hayward, Hoppes, and
  Hudson}]{PhysRev.105.1413}
\bibinfo{author}{C.~S. Wu}, \bibinfo{author}{E.~Ambler}, \bibinfo{author}{R.~W.
  Hayward}, \bibinfo{author}{D.~D. Hoppes}, \bibinfo{author}{R.~P. Hudson},
  \bibinfo{journal}{Phys. Rev.} \bibinfo{volume}{105} (\bibinfo{year}{1957})
  \bibinfo{pages}{1413--1415}. \DOIprefix\doi{10.1103/PhysRev.105.1413}.
\bibitem[{{Garcia Ruiz} et~al.(2015){Garcia Ruiz}, Bissell, Gottberg, Stachura,
  Hemmingsen, Neyens, and Severijns}]{rogar2015}
\bibinfo{author}{R.~F. {Garcia Ruiz}}, \bibinfo{author}{M.~L. Bissell},
  \bibinfo{author}{A.~Gottberg}, \bibinfo{author}{M.~Stachura},
  \bibinfo{author}{L.~Hemmingsen}, \bibinfo{author}{G.~Neyens},
  \bibinfo{author}{N.~Severijns}, \bibinfo{journal}{EPJ Web of Conferences}
  \bibinfo{volume}{93} (\bibinfo{year}{2015}) \bibinfo{pages}{07004}.
  \DOIprefix\doi{10.1051/epjconf/20159307004}.
\bibitem[{Kowalska et~al.(2018)Kowalska, Araujo~Escalona, Baranowski, Croese,
  Cerato, Bissell, Gins, Gustafsson, Harding, Hemmingsen, Heylen, Hofmann,
  Kanellakopoulos, Kocman, Kozak, Madurga~Flores, Neyens, Pallada, Plavec,
  Szutkowski, Walczak, Wienholtz, Wolak, Yang, and Zakoucky}]{Kowalska:2299798}
\bibinfo{author}{M.~Kowalska}, \bibinfo{author}{V.~Araujo~Escalona},
  \bibinfo{author}{M.~Baranowski}, \bibinfo{author}{J.~Croese},
  \bibinfo{author}{L.~Cerato}, \bibinfo{author}{M.~Bissell},
  \bibinfo{author}{W.~Gins}, \bibinfo{author}{F.~Gustafsson},
  \bibinfo{author}{R.~Harding}, \bibinfo{author}{L.~Hemmingsen},
  \bibinfo{author}{H.~Heylen}, \bibinfo{author}{F.~Hofmann},
  \bibinfo{author}{A.~Kanellakopoulos}, \bibinfo{author}{V.~Kocman},
  \bibinfo{author}{M.~Kozak}, \bibinfo{author}{M.~Madurga~Flores},
  \bibinfo{author}{G.~Neyens}, \bibinfo{author}{S.~Pallada},
  \bibinfo{author}{J.~Plavec}, \bibinfo{author}{K.~Szutkowski},
  \bibinfo{author}{M.~Walczak}, \bibinfo{author}{F.~Wienholtz},
  \bibinfo{author}{J.~Wolak}, \bibinfo{author}{X.~Yang},
  \bibinfo{author}{D.~Zakoucky}, \bibinfo{title}{{Interaction of Na+ ions with
  DNA G-quadruplex structures studied directly with Na beta-NMR spectroscopy}},
  \bibinfo{type}{Technical Report} \bibinfo{number}{CERN-INTC-2018-019.
  INTC-P-521-ADD-1}, CERN, \bibinfo{address}{Geneva}, \bibinfo{year}{2018}.
\bibitem[{Velten et~al.(2014)Velten, Bissell, Neyens, and
  Severijns}]{Velten2014}
\bibinfo{author}{P.~Velten}, \bibinfo{author}{M.~L. Bissell},
  \bibinfo{author}{G.~Neyens}, \bibinfo{author}{N.~Severijns},
  \bibinfo{title}{{Measurement of the $\beta$-asymmetry parameter in $^{35}$Ar
  decay with a laser polarized beam}}, \bibinfo{type}{Technical Report}
  \bibinfo{number}{CERN-INTC-2014-062. INTC-P-426}, CERN,
  \bibinfo{address}{Geneva}, \bibinfo{year}{2014}.
\bibitem[{Kastler(1957)}]{Kastler1957}
\bibinfo{author}{A.~Kastler}, \bibinfo{journal}{J. Opt. Soc. Am.}
  \bibinfo{volume}{47} (\bibinfo{year}{1957}) \bibinfo{pages}{460--465}.
  \DOIprefix\doi{10.1364/JOSA.47.000460}.
\bibitem[{Demtr{\"{o}}der(1981)}]{Demtroder1981}
\bibinfo{author}{W.~Demtr{\"{o}}der}, \bibinfo{title}{{Laser Spectroscopy -
  Basic Concepts and Instrumentation}}, \bibinfo{publisher}{Springer-Verlag},
  \bibinfo{year}{1981}.
\bibitem[{Neugart et~al.(2017)Neugart, Billowes, Bissell, Blaum, Cheal,
  Flanagan, Neyens, N{\"{o}}rtersh{\"{a}}user, and Yordanov}]{Neugart2017}
\bibinfo{author}{R.~Neugart}, \bibinfo{author}{J.~Billowes},
  \bibinfo{author}{M.~L. Bissell}, \bibinfo{author}{K.~Blaum},
  \bibinfo{author}{B.~Cheal}, \bibinfo{author}{K.~T. Flanagan},
  \bibinfo{author}{G.~Neyens}, \bibinfo{author}{W.~N{\"{o}}rtersh{\"{a}}user},
  \bibinfo{author}{D.~T. Yordanov}, \bibinfo{journal}{Journal of Physics G:
  Nuclear and Particle Physics} \bibinfo{volume}{44} (\bibinfo{year}{2017})
  \bibinfo{pages}{064002}. \DOIprefix\doi{10.1088/1361-6471/aa6642}.
\bibitem[{Jackson et~al.(1957)Jackson, Treiman, and Wyld}]{Jackson1957}
\bibinfo{author}{J.~D. Jackson}, \bibinfo{author}{S.~B. Treiman},
  \bibinfo{author}{H.~W. Wyld}, \bibinfo{journal}{Phys. Rev.}
  \bibinfo{volume}{106} (\bibinfo{year}{1957}) \bibinfo{pages}{517--521}.
  \DOIprefix\doi{10.1103/PhysRev.106.517}.
\bibitem[{Yordanov(2007)}]{Yordanov2007}
\bibinfo{author}{D.~T. Yordanov}, \bibinfo{title}{{From $^{27}$Mg to $^{33}$Mg:
  transition to the Island of inversion}}, Ph.D. thesis, Katholieke
  Universiteit Leuven, \bibinfo{year}{2007}.
\bibitem[{Gins(tion)}]{Gins2018b}
\bibinfo{author}{W.~Gins}, \bibinfo{title}{{Development of a dedicated
  laser-polarization beamline for ISOLDE-CERN}}, Ph.D. thesis, KULeuven,
  \bibinfo{year}{In preparation}.
\bibitem[{Stachura et~al.(2016)Stachura, Gottberg, Johnston, Bissell, {Garcia
  Ruiz}, {Martins Correia}, {Granadeiro Costa}, Dehn, Deicher, Fenta,
  Hemmingsen, M{\o}lholt, Munch, Neyens, Pallada, Silva, and
  Zakoucky}]{Stachura2016}
\bibinfo{author}{M.~Stachura}, \bibinfo{author}{A.~Gottberg},
  \bibinfo{author}{K.~Johnston}, \bibinfo{author}{M.~L. Bissell},
  \bibinfo{author}{R.~F. {Garcia Ruiz}}, \bibinfo{author}{J.~{Martins
  Correia}}, \bibinfo{author}{A.~R. {Granadeiro Costa}},
  \bibinfo{author}{M.~Dehn}, \bibinfo{author}{M.~Deicher},
  \bibinfo{author}{A.~Fenta}, \bibinfo{author}{L.~Hemmingsen},
  \bibinfo{author}{T.~E. M{\o}lholt}, \bibinfo{author}{M.~Munch},
  \bibinfo{author}{G.~Neyens}, \bibinfo{author}{S.~Pallada},
  \bibinfo{author}{M.~R. Silva}, \bibinfo{author}{D.~Zakoucky},
  \bibinfo{journal}{Nuclear Instruments and Methods in Physics Research,
  Section B: Beam Interactions with Materials and Atoms} \bibinfo{volume}{376}
  (\bibinfo{year}{2016}) \bibinfo{pages}{369--373}.
  \DOIprefix\doi{10.1016/j.nimb.2016.02.030}.
\bibitem[{Kugler(2000)}]{Kugler2000}
\bibinfo{author}{E.~Kugler}, \bibinfo{journal}{Hyperfine Interactions}
  \bibinfo{volume}{129} (\bibinfo{year}{2000}) \bibinfo{pages}{23--42}.
  \DOIprefix\doi{10.1023/A:1012603025802}.
\bibitem[{Kowalska et~al.(2008)Kowalska, Yordanov, Blaum, Himpe, Lievens,
  Mallion, Neugart, Neyens, and Vermeulen}]{Kowalska2008}
\bibinfo{author}{M.~Kowalska}, \bibinfo{author}{D.~T. Yordanov},
  \bibinfo{author}{K.~Blaum}, \bibinfo{author}{P.~Himpe},
  \bibinfo{author}{P.~Lievens}, \bibinfo{author}{S.~Mallion},
  \bibinfo{author}{R.~Neugart}, \bibinfo{author}{G.~Neyens},
  \bibinfo{author}{N.~Vermeulen}, \bibinfo{journal}{Physical Review C - Nuclear
  Physics} \bibinfo{volume}{77} (\bibinfo{year}{2008}) \bibinfo{pages}{1--11}.
  \DOIprefix\doi{10.1103/PhysRevC.77.034307}.
\bibitem[{Kreim et~al.(2014)Kreim, Bissell, Papuga, Blaum, Rydt, Ruiz, Goriely,
  Heylen, Kowalska, Neugart, Neyens, Nörtershäuser, Rajabali, Alarcón,
  Stroke, and Yordanov}]{kreim2014}
\bibinfo{author}{K.~Kreim}, \bibinfo{author}{M.~Bissell},
  \bibinfo{author}{J.~Papuga}, \bibinfo{author}{K.~Blaum},
  \bibinfo{author}{M.~D. Rydt}, \bibinfo{author}{R.~G. Ruiz},
  \bibinfo{author}{S.~Goriely}, \bibinfo{author}{H.~Heylen},
  \bibinfo{author}{M.~Kowalska}, \bibinfo{author}{R.~Neugart},
  \bibinfo{author}{G.~Neyens}, \bibinfo{author}{W.~Nörtershäuser},
  \bibinfo{author}{M.~Rajabali}, \bibinfo{author}{R.~S. Alarcón},
  \bibinfo{author}{H.~Stroke}, \bibinfo{author}{D.~Yordanov},
  \bibinfo{journal}{Physics Letters B} \bibinfo{volume}{731}
  (\bibinfo{year}{2014}) \bibinfo{pages}{97 -- 102}.
  \DOIprefix\doi{https://doi.org/10.1016/j.physletb.2014.02.012}.
\bibitem[{Foot(2005)}]{Foot2005a}
\bibinfo{author}{C.~J. Foot}, \bibinfo{title}{{Atomic Physics}}, Oxford Master
  Series in Physics, \bibinfo{publisher}{Oxford University Press},
  \bibinfo{year}{2005}.
\bibitem[{Garcia~Ruiz et~al.(2017)Garcia~Ruiz, Gorges, Bissell, Blaum, Gins,
  Heylen, Koenig, Kaufmann, Kowalska, Kramer, Lievens, Malbrunot-Ettenauer,
  Neugart, Neyens, N{\"{o}}rtershauser, Yordanov, and Yang}]{Ruiz2017}
\bibinfo{author}{R.~F. Garcia~Ruiz}, \bibinfo{author}{C.~Gorges},
  \bibinfo{author}{M.~Bissell}, \bibinfo{author}{K.~Blaum},
  \bibinfo{author}{W.~Gins}, \bibinfo{author}{H.~Heylen},
  \bibinfo{author}{K.~Koenig}, \bibinfo{author}{S.~Kaufmann},
  \bibinfo{author}{M.~Kowalska}, \bibinfo{author}{J.~Kramer},
  \bibinfo{author}{P.~Lievens}, \bibinfo{author}{S.~Malbrunot-Ettenauer},
  \bibinfo{author}{R.~Neugart}, \bibinfo{author}{G.~Neyens},
  \bibinfo{author}{W.~N{\"{o}}rtershauser}, \bibinfo{author}{D.~T. Yordanov},
  \bibinfo{author}{X.~F. Yang}, \bibinfo{journal}{Journal of Physics G: Nuclear
  and Particle Physics} \bibinfo{volume}{44} (\bibinfo{year}{2017}).
  \DOIprefix\doi{10.1088/1361-6471/aa5a24}.
\bibitem[{Man{\'{e}} et~al.(2009)Man{\'{e}}, Billowes, Blaum, Campbell, Cheal,
  Delahaye, Flanagan, Forest, Franberg, Geppert, Giles, Jokinen, Kowalska,
  Neugart, Neyens, N{\"{o}}rtersh{\"{a}}user, Podadera, Tungate, Vingerhoets,
  and Yordanov}]{Mane2009}
\bibinfo{author}{E.~Man{\'{e}}}, \bibinfo{author}{J.~Billowes},
  \bibinfo{author}{K.~Blaum}, \bibinfo{author}{P.~Campbell},
  \bibinfo{author}{B.~Cheal}, \bibinfo{author}{P.~Delahaye},
  \bibinfo{author}{K.~T. Flanagan}, \bibinfo{author}{D.~H. Forest},
  \bibinfo{author}{H.~Franberg}, \bibinfo{author}{C.~Geppert},
  \bibinfo{author}{T.~Giles}, \bibinfo{author}{A.~Jokinen},
  \bibinfo{author}{M.~Kowalska}, \bibinfo{author}{R.~Neugart},
  \bibinfo{author}{G.~Neyens}, \bibinfo{author}{W.~N{\"{o}}rtersh{\"{a}}user},
  \bibinfo{author}{I.~Podadera}, \bibinfo{author}{G.~Tungate},
  \bibinfo{author}{P.~Vingerhoets}, \bibinfo{author}{D.~T. Yordanov},
  \bibinfo{journal}{European Physical Journal A} \bibinfo{volume}{42}
  (\bibinfo{year}{2009}) \bibinfo{pages}{503--507}.
  \DOIprefix\doi{10.1140/epja/i2009-10828-0}.
\bibitem[{Floettmann(2003)}]{Floetmann2003}
\bibinfo{author}{K.~Floettmann}, \bibinfo{journal}{Phys. Rev. ST Accel. Beams}
  \bibinfo{volume}{6} (\bibinfo{year}{2003}) \bibinfo{pages}{034202}.
  \DOIprefix\doi{10.1103/PhysRevSTAB.6.034202}.
\bibitem[{{COMSOL}(2017)}]{comsol}
\bibinfo{author}{{COMSOL}}, \bibinfo{title}{Multiphysics Reference Guide for
  COMSOL 5.2a}, \bibinfo{year}{2017}.
\bibitem[{Keim(1996)}]{matthias1996}
\bibinfo{author}{M.~Keim}, \bibinfo{title}{Messung der Kernquadrupolmomente
  neutronenreicher Natriumisotope}, Ph.D. thesis, Uni Mainz,
  \bibinfo{year}{1996}.
\bibitem[{Levy et~al.(2010)Levy, Pearson, Morris, Chow, Hossain, Kiefl,
  Labb{\'e}, Lassen, MacFarlane, Parolin, Saadaoui, Smadella, Song, and
  Wang}]{Levy2010}
\bibinfo{author}{C.~D.~P. Levy}, \bibinfo{author}{M.~R. Pearson},
  \bibinfo{author}{G.~D. Morris}, \bibinfo{author}{K.~H. Chow},
  \bibinfo{author}{M.~D. Hossain}, \bibinfo{author}{R.~F. Kiefl},
  \bibinfo{author}{R.~Labb{\'e}}, \bibinfo{author}{J.~Lassen},
  \bibinfo{author}{W.~A. MacFarlane}, \bibinfo{author}{T.~J. Parolin},
  \bibinfo{author}{H.~Saadaoui}, \bibinfo{author}{M.~Smadella},
  \bibinfo{author}{Q.~Song}, \bibinfo{author}{D.~Wang},
  \bibinfo{journal}{Hyperfine Interactions} \bibinfo{volume}{196}
  (\bibinfo{year}{2010}) \bibinfo{pages}{287--294}.
  \DOIprefix\doi{10.1007/s10751-009-0148-9}.
\bibitem[{Welte et~al.(2010)Welte, Ritterbusch, Steinke, Henrich,
  Aeschbach-Hertig, and Oberthaler}]{Weite2010}
\bibinfo{author}{J.~Welte}, \bibinfo{author}{F.~Ritterbusch},
  \bibinfo{author}{I.~Steinke}, \bibinfo{author}{M.~Henrich},
  \bibinfo{author}{W.~Aeschbach-Hertig}, \bibinfo{author}{M.~K. Oberthaler},
  \bibinfo{journal}{New Journal of Physics} \bibinfo{volume}{12}
  (\bibinfo{year}{2010}). \DOIprefix\doi{10.1088/1367-2630/12/6/065031}.
\bibitem[{Mertzimekis et~al.(2016)Mertzimekis, Stamou, and
  Psaltis}]{2016_Mertzimekis}
\bibinfo{author}{T.~Mertzimekis}, \bibinfo{author}{K.~Stamou},
  \bibinfo{author}{A.~Psaltis}, \bibinfo{journal}{Nuclear Instruments and
  Methods in Physics Research Section A: Accelerators, Spectrometers, Detectors
  and Associated Equipment} \bibinfo{volume}{807} (\bibinfo{year}{2016})
  \bibinfo{pages}{56--60}.
  \DOIprefix\doi{http://dx.doi.org/10.1016/j.nima.2015.10.096}.
\bibitem[{Gins et~al.(2018)Gins, de~Groote, Bissell, {Granados Buitrago},
  Ferrer, Lynch, Neyens, and Sels}]{Gins2018}
\bibinfo{author}{W.~Gins}, \bibinfo{author}{R.~P. de~Groote},
  \bibinfo{author}{M.~L. Bissell}, \bibinfo{author}{C.~{Granados Buitrago}},
  \bibinfo{author}{R.~Ferrer}, \bibinfo{author}{K.~M. Lynch},
  \bibinfo{author}{G.~Neyens}, \bibinfo{author}{S.~Sels},
  \bibinfo{journal}{Computer Physics Communications} \bibinfo{volume}{222}
  (\bibinfo{year}{2018}) \bibinfo{pages}{286--294}.
  \DOIprefix\doi{10.1016/j.cpc.2017.09.012}.
\bibitem[{Johansson et~al.(2013)Johansson, Nation, and Nori}]{johansson2013}
\bibinfo{author}{J.~Johansson}, \bibinfo{author}{P.~Nation},
  \bibinfo{author}{F.~Nori}, \bibinfo{journal}{Computer Physics Communications}
  \bibinfo{volume}{184} (\bibinfo{year}{2013}) \bibinfo{pages}{1234 -- 1240}.
  \DOIprefix\doi{https://doi.org/10.1016/j.cpc.2012.11.019}.
\bibitem[{Touchard et~al.(1982)Touchard, Serre, B\"uttgenbach, Guimbal,
  Klapisch, de~Saint~Simon, Thibault, Duong, Juncar, Liberman, Pinard, and
  Vialle}]{Touchard1982}
\bibinfo{author}{F.~Touchard}, \bibinfo{author}{J.~M. Serre},
  \bibinfo{author}{S.~B\"uttgenbach}, \bibinfo{author}{P.~Guimbal},
  \bibinfo{author}{R.~Klapisch}, \bibinfo{author}{M.~de~Saint~Simon},
  \bibinfo{author}{C.~Thibault}, \bibinfo{author}{H.~T. Duong},
  \bibinfo{author}{P.~Juncar}, \bibinfo{author}{S.~Liberman},
  \bibinfo{author}{J.~Pinard}, \bibinfo{author}{J.~L. Vialle},
  \bibinfo{journal}{Phys. Rev. C} \bibinfo{volume}{25} (\bibinfo{year}{1982})
  \bibinfo{pages}{2756--2770}. \DOIprefix\doi{10.1103/PhysRevC.25.2756}.

\end{thebibliography}
